\documentclass[conference]{IEEEtran}
\IEEEoverridecommandlockouts
\usepackage{cite}
\usepackage{amsmath,amssymb,amsfonts}
\usepackage{algorithmic}
\usepackage{graphicx}
\usepackage{textcomp}
\usepackage{xcolor}
\usepackage{bbding}

\usepackage{color, colortbl}
\definecolor{Gray}{gray}{0.95}

\def\BibTeX{{\rm B\kern-.05em{\sc i\kern-.025em b}\kern-.08em
    T\kern-.1667em\lower.7ex\hbox{E}\kern-.125emX}}
    
\begin{document}

\title{EMA2S: An End-to-End Multimodal Articulatory-to-Speech System}

\author{\IEEEauthorblockN{Yu-Wen Chen$^1$, Kuo-Hsuan Hung$^1$, Shang-Yi Chuang$^1$,\\ Jonathan Sherman$^1$, Wen-Chin Huang$^2{^3}$, Xugang Lu$^4$, Yu Tsao$^1$}
\IEEEauthorblockA{$^1$Research Center for Information Technology Innovation, Academia Sinica, Taiwan}
\IEEEauthorblockA{$^2$Institute of Information Science, Academia Sinica, Taiwan}
\IEEEauthorblockA{$^3$Graduate School of Informatics, Nagoya University, Japan}
\IEEEauthorblockA{$^4$National Institute of Information and Communications Technology, Japan}
}

\maketitle

\begin{abstract}

Synthesized speech from articulatory movements can have real-world use for patients with vocal cord disorders, situations requiring silent speech, or in high-noise environments. In this work, we present EMA2S, an end-to-end multimodal articulatory-to-speech system that directly converts articulatory movements to speech signals. We use a neural-network-based vocoder combined with multimodal joint-training, incorporating spectrogram, mel-spectrogram, and deep features. The experimental results confirm that the multimodal approach of EMA2S outperforms the baseline system in terms of both objective evaluation and subjective evaluation metrics. Moreover, results demonstrate that joint mel-spectrogram and deep feature loss training can effectively improve system performance.

\end{abstract}


\section{Introduction}

Silent speech interfaces enable people to communicate without the presence of an acoustic signal. Such techniques can provide patients who suffer from vocal cord disorders a more natural alternative way to communicate \cite{lin2020end}. Also, these techniques can be helpful in situations requiring acoustic silence, or in high-noise environments, since acoustic signals are not required as input and thus background noises have a greatly reduced effect. Various silent speech technologies have been investigated, including magnetic resonance imaging \cite{badin2002three, rathinavelu2007three}, electromyograms \cite{janke2017emg}, permanent magnetic articulograph \cite{cao2019permanent}, and electromagnetic midsagittal articulography (EMA) \cite{rudzicz2010learning, wang2012phoneme, rudzicz2012torgo, li2012cross}.

In this study, we use EMA to collect the articulatory movements data. EMA records the articulatory movements by using an electromagnetic field to induce currents in sensors, which are attached to articulators such as lips and tongue. Previous studies have proposed several methods to convert EMA signals towards acoustic features. \cite{kaburagi1998determination} uses a codebook to store articulatory and acoustic parameters pairs, and then estimates the spectrum of the articulatory data by selecting neighbor samples in the codebook. Also, statistical models such as Gaussian mixture models (GMM) \cite{toda2004mapping}, hidden Markov models (HMM) \cite{hueber2016statistical}, fully connected neural network \cite{aryal2016data}, and bidirectional long short-term memory (BLSTM) \cite{liu2016articulatory, taguchi2018articulatory} have been used to map the articulatory movements to acoustic signals. These studies have indicated that neural-network-based methods achieve better performance than GMM and HMM methods. However, they only use neural networks to map the articulatory movements to spectral features, and reconstruct the waveform with traditional parametric vocoders such as STRAIGHT \cite{kawahara1999restructuring} and WORLD \cite{morise2016world}. Since the neural-network-based vocoders \cite{oord2016wavenet, yamamoto2020parallel} have shown much superior performance over traditional parametric vocoders, it is logical to investigate the performance of using neural networks for both articulatory-to-spectrum mapping and waveform reconstruction.
We propose an end-to-end multimodal articulatory-to-speech system, EMA2S, that improves the existing speech synthesis systems by applying two techniques: (1) a neural-network-based vocoder and (2) a multimodal jointly training method with a combined loss. Concerning the first, in addition to demonstrating much superior performance over the traditional parametric vocoders, the neural-network-based vocoder allows further development in an end-to-end trainable system that directly converts articulatory movements into waveforms. It is not bounded or required to fit the constraints of parametric or independently trained vocoders. For the second, we jointly train with a combined loss of different acoustic features: (a) the spectrogram loss, (b) the mel-spectrogram loss, and (c) the deep feature loss of spectral embeddings and articulatory movement embeddings. The deep feature loss \cite{germain2019speech} measures the dissimilarity between articulatory movement embeddings and spectral embeddings. To calculate the deep feature loss, both articulatory movements and spectrograms are used as input data during training, but only articulatory movements are necessary during inference. The introduction of the deep feature loss allows synthesis models to learn a better representation of one modality (articulatory movements) from multiple modalities (spectrograms and articulatory movements). 

Experimental results show that our proposed system outperforms a previous system in terms of mel-cepstral distortion (MCD) \cite{kubichek1993mel}, perceptual evaluation of speech quality (PESQ) \cite{rix2001perceptual}, short-time objective intelligibility (STOI) \cite{taal2011algorithm}, character correct rate (CCR) of a pre-trained automatic speech recognition (ASR) system \cite{ref_google_asr}, and a listening test. For the reason that users will be more willing to use the device without using invasive sensors, we investigate the system performance with only four less invasive EMA sensors. The results reveal that our proposed system still performs better than the previous system. 

The rest of the paper is organized as follows. Section \ref{related_work} introduces the related works. The proposed EMA2S system is presented in Section \ref{proposed_method}. Experimental details and results are given in Section \ref{experiments} to demonstrate the performance of the proposed approach. Section \ref{conclusion} concludes our work.

\section{Related Work \label{related_work}}
In this section, we review Parallel WaveGAN (PWG) \cite{yamamoto2020parallel}, multimodal learning \cite{ngiam2011multimodal}, and deep feature loss \cite{germain2019speech} used in our proposed model.

\subsection{Parallel WaveGAN}

PWG \cite{yamamoto2020parallel} is a non-autoregressive, fast, and effective parallel waveform generation method based on a generative adversarial network \cite{goodfellow2014generative}. PWG has shown superior performance to parametric vocoders, and can train and inference faster than autoregressive generative models such as WaveNet \cite{oord2016wavenet}. 

PWG uses a joint training method of the multi-resolution short-time Fourier transform (STFT) loss and the waveform-domain adversarial loss. To calculate the adversarial loss, PWG is composed of two separate neural networks: a generator and a discriminator. The input of the generator is auxiliary acoustic features, which are mel-spectrograms and random noises drawn from a Gaussian distribution, and the output of the generator is the raw waveform in parallel. The generator learns a distribution of realistic waveforms by trying to deceive the discriminator to classify the generated samples as real. On the contrary, the discriminator learns by correctly recognizing the generated sample as fake and the ground truth sample as real. 

\subsection{Multimodal Learning}

Multimodal learning \cite{ngiam2011multimodal} aims to learn relating information from multiple modalities and fill the missing modality given the observed ones. Numerous research has investigated the effectiveness of incorporating different features into speech-related systems, including text \cite{kinoshita2015text, huang2019voice, wang2020end, zhang2020transfer}, videos \cite{michelsanti2019deep, chuang2020lite, hou2018audio}, bone-conducted microphone signals \cite{yu2020time}, electropalatography \cite{chen2021multi}, and articulatory movements \cite{steiner2017synthesis, li2017multimodal, chen2020study}.

\subsection{Deep Feature Loss}

Deep feature loss \cite{germain2019speech} is defined as the dissimilarity of the embeddings in neural networks. Previous research has shown that deep features can capture the perceptual features of the input, and deep feature loss can effectively improve the model performance without adding the complexity of the processing network itself \cite{gatys2015texture, hou2017deep, germain2019speech}. 

\section{Proposed Method \label{proposed_method}}

In this work, we propose an end-to-end multimodal articulatory-to-speech system (EMA2S) that uses PWG \cite{yamamoto2020parallel} as a vocoder and incorporates multimodal learning \cite{ngiam2011multimodal} and deep feature loss \cite{germain2019speech}. Unlike previous studies, the deep feature loss in this work exploits the idea of multimodal learning, and calculates the dissimilarity of two modalities' embeddings (EMA embeddings and spectral embeddings) instead of one (EMA embeddings). The combination of multimodal learning and deep feature loss is designed for low resource data such as EMA signals since a network that extracts deep features of low resource data (e.g., EMA signals) is more difficult to obtain or train than a network that extracts deep features of high resource data (e.g., audio signals). Furthermore, given the objective to transform EMA signals to speech, the deep feature loss calculated by EMA embeddings and spectral embeddings aligns the training of the system.

\subsection{Architecture}

\begin{figure*}
\centerline{\includegraphics[scale=0.9]{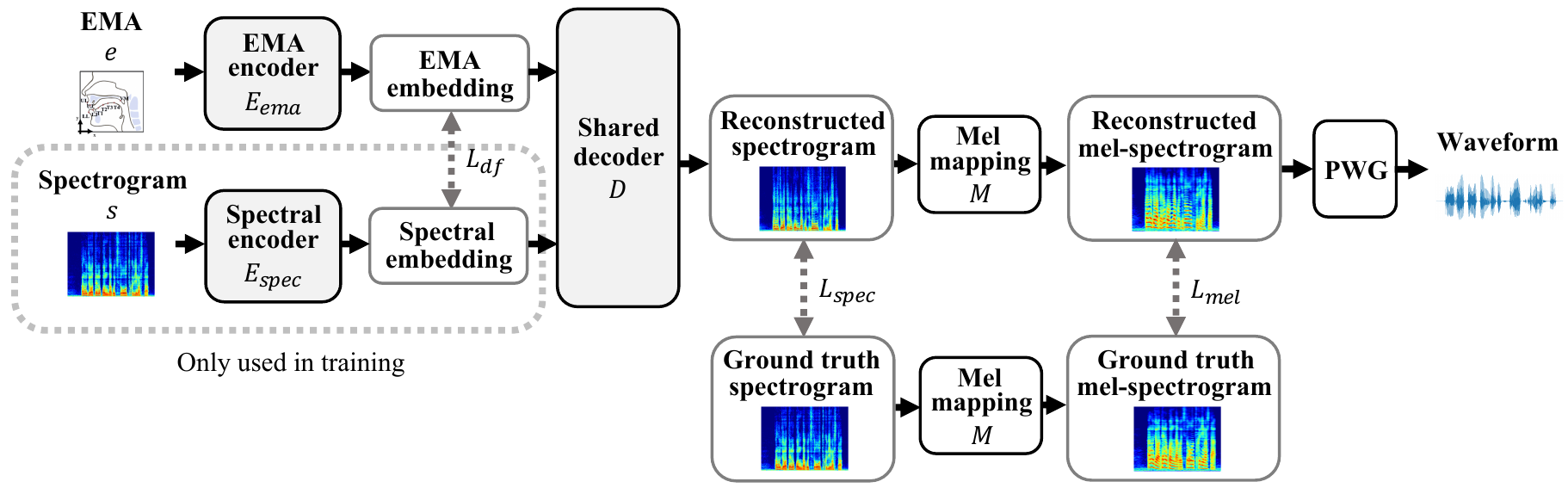}}
\caption{Diagram of EMA2S system.}
\label{architecture}
\end{figure*}

Fig.~\ref{architecture} depicts the proposed system, which includes an EMA encoder $E_{ema}$, a spectral encoder $E_{spec}$, and a shared decoder $D$. The spectral encoder is for guiding the training process of the EMA encoder and the shared decoder, and the spectral encoder will not be used in inference - only EMA features are required in testing. The output of the decoder is a spectrogram. This spectrogram will be transformed into a mel-spectrogram and then reconstructed back to waveform by the PWG model.

The spectral encoder contains two BLSTM layers, and the hidden units of the first and second layers are 196 and 256, respectively. The BLSTM layers are followed by a linear layer with 256 units and the rectified linear unit (ReLU). The EMA encoder consists of two BLSTM layers with 128 and 256 hidden units, followed by a linear layer with 256 units and the ReLU. The shared decoder consists of three BLSTM layers with 256 hidden units, a linear layer with 513 units, and the ReLU.

\subsection{Training Stages and Loss Function}

The training process contains two stages. The first stage is to train the spectral encoder $E_{spec}$ and the shared decoder $D$. The second stage is to train the EMA encoder $E_{ema}$ and shared decoder $D$.

In the first stage, $E_{spec}$ and $D$ are optimized by minimizing $L^{(1)}$ which is the reconstructed loss of spectral features, including spectrogram loss $L_{spec}^{(1)}$ and mel-spectrogram loss $L_{mel}^{(1)}$. $L^{(1)}$ is defined as follows:

\begin{equation}
\renewcommand\arraystretch{1.2}
\begin{aligned}
L_{spec}^{(1)} &=|D(E_{spec}(s))-s|\\
L_{mel}^{(1)} &=|M(D(E_{spec}(s)))-M(s)|\\
L^{(1)}       &= L_{spec}^{(1)} + L_{mel}^{(1)}
\end{aligned}
\end{equation}

{\noindent}where $s$ is the input spectrogram and $M$ is the mapping from spectrogram to mel-spectrogram.

In the second stage, $E_{ema}$ and $D$ are optimized by minimizing $L^{(2)}$ which combines the reconstructed spectrogram loss $L_{spec}^{(2)}$, the reconstructed mel-spectrogram loss $L_{mel}^{(2)}$, and the deep feature loss $L_{df}^{(2)}$. In this stage, the spectrogram and mel-spectrogram are reconstructed from EMA embeddings rather than spectral embeddings. The deep feature loss $L_{df}^{(2)}$ measures the dissimilarity between EMA embeddings and spectral embeddings. We want the EMA embeddings close to spectral embeddings because we assume that we can more easily reconstruct the spectrograms and mel-spectrograms by the spectral embeddings. $L^{(2)}$ is defined as follows: 

\begin{equation}
\renewcommand\arraystretch{1.2}
\begin{aligned}
L_{spec}^{(2)} &=|D(E_{ema}(e))-s|\\
L_{mel}^{(2)} &=|M(D(E_{ema}(e)))-M(s)|\\
L_{df}^{(2)} &= |E_{ema}(e)-E_{spec}(s)|\\
L^{(2)}       &= L_{spec}^{(2)} + L_{mel}^{(2)}+L_{df}^{(2)}
\end{aligned}
\end{equation}

{\noindent}where $e$ is the input EMA signal.

\section{Experiments \label{experiments}}

\subsection{Experimental setup}

\begin{figure}[htbp]
\centerline{\includegraphics[scale=0.8]{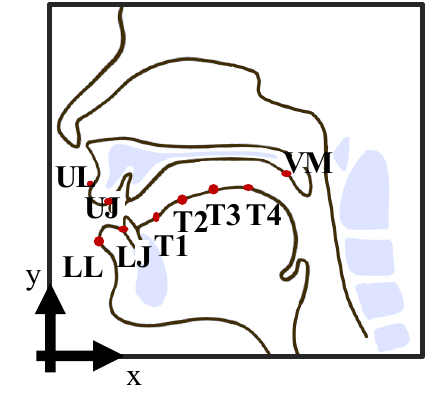}}
\caption{The placement of EMA sensors.}
\label{EMA}
\end{figure}

In this study, we use the EMA data collected by NTT, Tokyo, Japan \cite{okadome2001generation}. The sensor coils of EMA sensors are placed at the upper lip (UL), lower lip (LL), upper jaw (UJ), lower jaw (LJ), tongue tip (T1), tongue blade (T2), tongue dorsum (T3), tongue rear (T4), and velum (VM) as shown in Fig.~\ref{EMA}. EMA records the Cartesian coordinates of each sensor point at a sampling rate of 250 Hz, and the audio signals are recorded at the same time with a sampling rate of 16 kHz. The dataset contains articulatory movements and speech signals from three speakers, each providing 354 utterances. The training set includes 304 utterances from each speaker, and the testing set includes the remaining 50 utterances.

We divide the EMA signals by the maximum value for normalization, and we concatenate $\pm$ frames for a five-frame context window. For speech signals, we convert waveforms to spectrograms by STFT and only use the magnitude components to train the proposed model. The STFT settings are the same as that of the pre-trained PWG model, which use a window size of 1024 and a hop length of 256. The PWG used in this work was pre-trained under the Japanese corpus JNAS \cite{itou1999jnas}, which has the same sampling rate as that of our dataset.

We evaluate our results with objective evaluation metrics including MCD \cite{kubichek1993mel}, PESQ \cite{rix2001perceptual}, STOI \cite{taal2011algorithm}, and CCR of a pre-trained ASR \cite{ref_google_asr} system. We measure speech quality by using MCD and PESQ, and we evaluate speech intelligibility with STOI and CCR of a pre-trained ASR system. The CCR is calculated using Levenshtein distance \cite{levenshtein1966binary}.

For subjective evaluation, we conduct an A/B test for subjective listening tests. The A/B test compares the baseline system and the proposed system to determine which one brings better signal quality. The testing data contain five questions for each of the three speakers, resulting in a total of 15 questions.


\subsection{Baseline System}

The baseline system is based on the work in \cite{taguchi2018articulatory}, which is composed of three fully-connected layers, a layer normalization \cite{ba2016layer} layer, a sigmoid layer with 128 units, two layers of BLSTM with 256 units, and a fully-connected output layer. The input of the model is the EMA signal, while the target is the concatenated feature of mel-cepstrum, aperiodic parameters, F0 and voice activity detection (VAD). WORLD \cite{morise2016world} is used to extract the feature parameters of spectral envelope, aperiodic parameters and F0. The spectral envelope is further processed into mel-cepstrum, and F0 is further processed into VAD. Each feature parameter was normalized to zero mean and unit standard deviation. During inference, WORLD generates a speech signal using the synthesized speech parameters. Note that we skip the dynamic features (delta features) and maximum likelihood parameter generation algorithm used in \cite{taguchi2018articulatory} because we have already considered the forward and backward time series of the input EMA data.

\begin{table*}
\renewcommand\arraystretch{1.2}
\begin{center}
\begin{tabular}{|c|c|c|c|c|c|c|}
\hline
& Loss  & $E_{spec}$ &  MCD    & PESQ   & STOI & CCR  \\ \hline
\rowcolor{Gray} Baseline & - & - & 7.815 & 1.279 & 0.696 &0.818 \\ \hline
S{\tiny I}   & $L^{(2)}_{spec}$                 & \XSolidBrush   &  8.264  & 1.259          & 0.679          & 0.796         \\ \hline
S{\tiny II}  & $L^{(2)}_{spec}, L^{(2)}_{mel}$  & \XSolidBrush     &  7.334  & 1.320          & 0.702          & 0.841        \\ \hline
S{\tiny III}  & $L^{(1)}_{spec}, L^{(2)}_{spec},  L^{(2)}_{df}$ & \Checkmark  &  8.445  & 1.303          & 0.697          & 0.831 \\ \hline
EMA2S                                        & $L^{(1)}, L^{(2)}$               & \Checkmark   &  \textbf{7.176}  & \textbf{1.350} & \textbf{0.716} & \textbf{0.868} \\ \hline
\end{tabular}
\end{center}
\caption{Training loss and the numerical analysis of the articulatory-to-speech systems.}
\label{table:numerical_analysis}
\vspace{-1.8em}
\end{table*}

\subsection{Experimental results}

\subsubsection{Perceptual Analysis}

\begin{figure}[htbp]
\centerline{\includegraphics[scale=0.9]{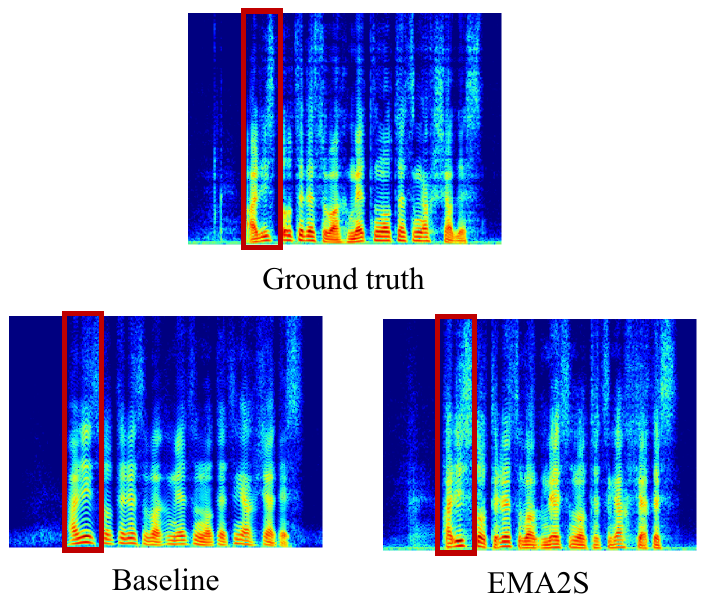}}
\caption{Visualization of spectrograms.}
\label{spec}
\end{figure}

Fig.~\ref{spec} visualizes the ground truth spectrograms as well as the reconstructed spectrograms. The results show that reconstructed spectrograms are visually close to the ground truths, which reveal that EMA signals can be successfully transformed into speech signals. Also, as indicated in the red boxes, the proposed EMA2S generated speech with more details in the high-frequency bands and less unnatural formant structures than the baseline system.

\subsubsection{Numerical Analysis}

To evaluate the combined loss, we compared EMA2S performance with three configurations of the articulatory-to-speech system (S{\footnotesize I}, S{\footnotesize II}, and S{\footnotesize III}). S{\footnotesize I} has no spectral encoder $E_{spec}$, and is trained with only spectrogram loss $L^{(2)}_{spec}$. S{\footnotesize II} is S{\footnotesize I} trained with both spectrogram loss $L^{(2)}_{spec}$ and mel-spectrogram loss $L^{(2)}_{mel}$. S{\footnotesize III} is S{\footnotesize I} with multimodal jointly training. It has a spectral encoder, and is trained with spectrogram loss $L^{(1)}_{spec}$, $L^{(2)}_{spec}$, and deep feature loss $L^{(2)}_{df}$.

Table \ref{table:numerical_analysis} organizes the corresponding training losses and shows the performance of the different articulatory-to-speech systems, including the baseline, S{\footnotesize I}, S{\footnotesize II}, S{\footnotesize III}, and the proposed EMA2S. The check mark in the column $E_{spec}$ indicates whether the system contains a spectral encoder that used for multimodal learning. The results reveal that EMA2S outperforms the baseline system in terms of MCD, PESQ, STOI, and CCR. Moreover, the performance of the articulatory-to-speech system can be improved by training the system with a combined loss of spectrogram and mel-spectrogram and using the multimodal jointly training method. 

\subsubsection{Listening Test}

\begin{figure}[htbp]
\centerline{\includegraphics[scale=0.91]{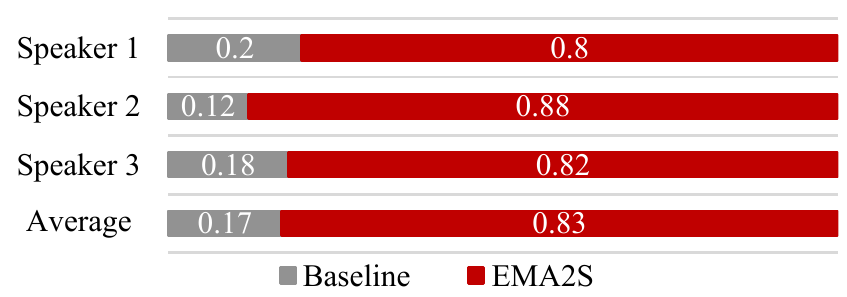}}
\caption{The results (in percentage) of the A/B listening test.}
\label{listening}
\end{figure}

We recruited 10 participants for an A/B test. Each participant must answer 15 questions, and each question contains two speech waveforms of the same utterance. One of the waveforms is generated by the baseline system, and the other is generated by our EMA2S system. The participants are asked to choose which waveform they prefer. Experimental results in Fig.~\ref{listening} reveals that an average of 83\% participants voted for the proposed EMA2S system, and the remaining 17\% of participants voted for the baseline system. 

\subsubsection{Further Analysis}

\begin{figure}[htbp]
\centerline{\includegraphics[scale=0.91]{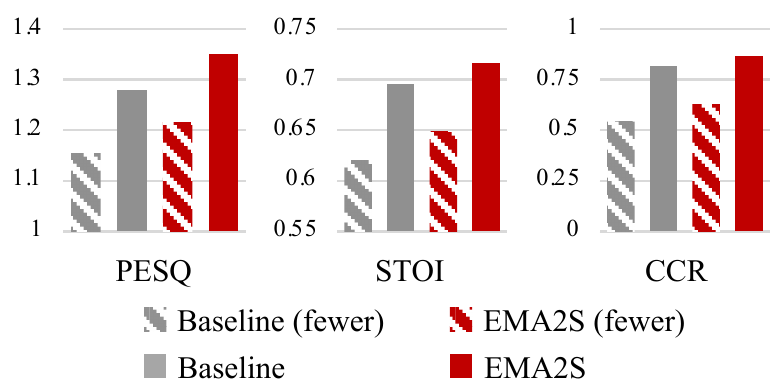}}
\caption{The average scores of different articulatory-to-speech systems.}
\label{fewer_sensor}
\end{figure}

Because EMA requires a laboratory environment for recording, we test EMA2S with only four less invasive sensors (UL, LL, LJ, and T1) to improve the applicability of the system, reasoning that without using invasive sensors, users will be more willing to use the devices. Fig.~\ref{fewer_sensor} shows that EMA2S with only four sensors (denoted as fewer) can still achieve better performance than the baseline system.

\section{Conclusion \label{conclusion}}

We propose EMA2S, an end-to-end multimodal articulatory-to-speech system that uses (1) a neural-network-based vocoder and (2) a multimodal jointly training method with a combined loss of spectrogram, mel-spectrogram, and the deep feature. Experimental results reveal that our proposed EMA2S system outperforms the baseline system in terms of objective evaluation metrics and a subjective listening test. In the future, we plan to increase the naturalness of the synthesized speech by incorporating a natural language model in the articulatory-to-speech system, and improve the performance of the system with limited sensors as input.


\section*{Acknowledgment}

We thank NTT Communication Science Laboratories for permitting us to use the articulatory data.

\bibliographystyle{IEEEtran}
\bibliography{refs}

\end{document}